\documentstyle[12pt]{article}

\begin{document}

\hoffset = -1truecm
\voffset = -2truecm

\title{\bf A new approach to parton recombination
in a QCD evolution equation }
\author{
{\bf
Wei Zhu
}\\
\normalsize Department of Physics, East China Normal University,
\normalsize Shanghai 200062, {\bf P.R. China}
}

\date{}

\newpage

\maketitle

\vskip 3truecm

\begin{abstract}

	Parton recombination is reconsidered in perturbation
theory without using the AGK cutting rules in the leading order of the
recombination. We use time-ordered perturbation theory  
to sum the cut diagrams, which are neglected in the GLR 
evolution equation. We present a set of new evolution equations including
parton recombination.

\end{abstract}

\newpage

\begin{center}
\section{Introduction}
\end{center}

	Parton recombination as a new higher twist phenomenon
was first discussed in the QCD evolution process by Gribov, Levin, Ryskin [1]
and Mueller, Qiu [2] in their pioneering works. This evolution equation
is called the GLR equation.

	An interesting effect of parton recombination is
screening or shadowing. In the case of higher number densities
of partons, for example in the small $x$ region, the gluons can overlap
spatially and annihilate. Therefore, one expects that the growth of the
gluon density with $Q^2$ will be suppressed by gluon recombination. These
suppression factors from the negative contributions due to gluon
recombination are calculated in the GLR equation
using the AGK (Abramovsky, Gribov, Kancheli) cutting
rules [3]. Assuming that the AGK cutting rules are valid in deep
inelastic scattering (DIS) in the small $x$ region, one finds that the
relative weights of cuts through two, one and zero ladders are $2:-4:1$,
as illustrated in fig. 1.  Due to these quantitative predictions of 
the suppression of parton number densities at small $x$, the GLR 
equation was extensively used to explore the structure of the nucleon 
and new perturbative QCD (PQCD) effects in the past years.

	However, the applications of AGK cutting rules in
the GLR equation have some drawbacks. For example, the
cut lines break the correlation between         
the recombining partons according to the AGK cutting rules in the GLR 
equation (fig. 1). As we will show in this work, the 
correlation among the initial partons in the QCD recombination equation
should be preserved. On the other hand, two-to-two parton processes 
may be associated with IR-divergences just as are the one-to-two 
processes in the Dokshitzer-Gribove- Lipatov-Altarelli-Parisi (DGLAP) 
evolution equation [4-6]. Although the AGK cutting rules provide
simple relations between the cross sections of hadron-hadron interactions
for different types of reggeon cuts, however,
the sum of cut graphs according to the AGK cutting rules cannot cancel
IR-divergences. The reason is that the difference of the contributions
between the positive graph and the negative graph is only a weight,
according to the AGK cutting rules. As we know that the virtual diagrams
are necessary for cancellation of IR-divergences in the DGLAP equation.
We will show in this work that above mentioned IR-divergences in two-to-two
processes also can be canceled by the sum of virtual diagrams, which are  
neglected in the GLR equation.

	In this work, we reconsider parton recombination 
in the QCD evolution equation without the AGK cutting rules.
To this end, we first point out that a new scale (the recombination
scale) exists in the parton recombination processes. We will give 
a definition of the recombination order of the process.
Then, we propose the bare probe-vertex
approximation. We find that several more diagrams, which
are neglected in the GLR equation, should be included in the QCD equation of
the parton fusion. We try to find a simple way to calculate
those cut diagrams in time-ordered perturbation theory (TOPT). 
Through a new derivation of the DGLAP evolution
equation, we present simple connections among the different 
cut diagrams. We shall show that both the shadowing-antishadowing 
and momentum conservation are the natural results of the theory.
As an interesting result, our new equation has different structure from  
the GLR equation. 

	The outline of the paper is the following. In section 2 we give
some definitions related to the parton model. In section 3 the sum of the
cut diagrams at the bare probe-vertex approximation is proposed. In
section 4 we give a new derivation of the DGLAP equation. Through
this example, we try to show the connections
among the relative cut graphs at the leading recombination order. 
The new evolution equations incorporating the parton
recombination are derived in sections  5--7. Section 8 contains the
discussions and concluding remarks.

\begin{center}
\section{Definitions}
\end{center}

	According to PQCD, a parton can always independently split into
two partons. However, except in the PQCD dynamics, the recombination
of partons depends on the overlap probability of their wave functions.
Therefore, we need a new physical quality to characterize parton 
recombination.

For example, consider an amplitude with two initial partons.
According to dimensional analysis, the hadronic part of the amplitude
should contain a factor $\sim 1/R$, where $R$ has dimension of length.
Now we incorporate the factor $1/Q$ arising from the partonic part
of the amplitude to form a dimensionless quality $1/(RQ)$. We call
this the recombination factor. For example, the recombination factor of
the process with two initial correlated partons is $1/(RQ)^2$. 
One can schematically think of $1/(RQ)^2$ as the overlap probability of 
two partons, where $1/Q$ is the scale of a parton at momentum 
transfer $Q^2$ and $R$ is the maximum correlation length of two partons. 
Usually, $R$ is regarded as the scale of target or the scale of 
the ``hot spots'', if they exist in the proton. This definition can 
be generalized to the case of the amplitudes including $m$-initial partons; 
the recombination factor in this case is $1/(RQ)^{m-1}$ if the fusing 
partons are paired.

	In this paper, we consider only the recombination processes at the
leading order level, that is, at $1/(RQ)^2$ and $\alpha_s^2$. Therefore,
we choose to study the following basic amplitudes as shown in fig. 2:
(a) $M^{(2-1)}_{p\gamma^*\rightarrow k'l'X}$,
(b) $M^{(2-2)}_{p\gamma^*\rightarrow k'l'X}$ and
(c) $M^{(2-3)}_{p\gamma^*\rightarrow k'l'X}$,
with the recombination factors
$1/(RQ)^0$, $1/(RQ)^1$ and $1/(RQ)^2$, respectively.
In fig. 2 we have omitted the distinction of the parton flavors; 
the dark circles indicate QCD interactions among the correlating partons.

	We see that the amplitudes involving parton recombination contain
the double scales:
$1/(RQ)$ and $\alpha_s$. We shall perform the calculations at a given
order-$1/(RQ)^m$ and order-$(\alpha_s)^n$ in two steps. First, we
calculate the process at the order of $1/(RQ)^2$. In this case,
the dark circles in fig. 2 are regarded as the elemental sub-processes.
The reason is that the decomposition of the circle part will break
the parton correlation and reduce the order of the recombination.
The second step is to calculate the sub-processes at order-$\alpha_s^2$
in PQCD. 

	We will use TOPT in this work.
Usually, TOPT is equivalent to the standard covariant perturbation
theory [7].  We shall call this TOPT as the normal TOPT (NTOPT),
where the time lines divide every basic vertex along the time-order.
In TOPT, the internal lines and the virtual particles are expressed by
external lines and effective real particles respectively. Therefore,
TOPT can also be used to describe amplitudes involving complex
vertices, where the part between two neighboring time lines can contain
a complex vertex. We define such a TOPT as an anomalous TOPT (ATOPT).
Obviously, ATOPT is not equivalent to NTOPT: they have different energy
deficits. On the other hand, there are energy-momentum
correlation between two neighboring complex vertices in ATOPT; therefore,
the vertex in ATOPT is not really factorized.

	We take the physical axial gauge, where the light-like vector
$n$ fixes the gauge as $n\cdot A=0$, $A$ being the gluon field.
The parton number densities are defined within the parton model
description of the photon nucleon DIS (fig. 3a) as
	
$$d\sigma(\gamma^*p\rightarrow k'X)=\int dx_1q(x_1)d\sigma(\gamma^*k
\rightarrow k'), \eqno (2.1)$$

where $q(x_1)dx_1$ is the number of quarks carrying momentum fraction
between $x_1$ and $x_1+dx_1$, where $x_1=k\cdot n/p\cdot n$. Formula (2.1)
means that the interaction of a virtual photon with proton can be
factorized as the soft part $q(x_1)$ and hard part
$d\sigma(\gamma^*k\rightarrow k')$. According to the parton model,

$$d\sigma(\gamma^*k\rightarrow k')=C_q\delta(x_1-x_B), \eqno (2.2)$$

where $x_B=Q^2/2p.q$ and $C_q$ is the coefficient depending on $x_B$ 
and $Q^2$. The quark density can be defined as

$$q(x_B)=\frac{1}{C_q}d\sigma(\gamma^*p\rightarrow k'X).\eqno(2.3)$$

On the other hand, using TOPT in the cut graph 3b, we have

$$d\sigma(\gamma^*p\rightarrow k'X)$$
$$=\frac{E_k}{E_p}\vert M_{p\rightarrow kX}\vert^2[\frac{1}{E_p-E_k-E_X}]^2
[\frac{1}{2E_k}]^2
\prod_X \frac{d^3k_X}{(2\pi)^32E_X}$$
$$\frac{1}{8E_kE_{\gamma}}\vert M_{{\gamma^*}k\rightarrow k'}\vert^2
(2\pi)^4\delta^4(p_\gamma+k-k')\frac{d^3k'}{(2\pi)^32E_{k'}}. \eqno(2.4)$$

Comparing eqs. (2.1) with (2.4), we get the definitions of the quark number
density

$$q(x_1)dx_1$$
$$=\frac{E_k}{E_p}\vert M_{p\rightarrow kX}\vert^2[\frac{1}{E_p-E_k-E_x}]^2
[\frac{1}{2E_k}]^2
\prod_X \frac{d^3k_X}{(2\pi)^32E_X}. \eqno(2.5)$$

and the bare probe-parton vertex

$$\frac{1}{C_q}d\sigma(\gamma^*k\rightarrow k')$$
$$=\frac{1}{C_q}\frac{1}{8E_kE_{\gamma}}\vert M_{{\gamma^*}k\rightarrow k'}\vert^2
(2\pi)^4\delta^4(p_\gamma+k-k')\frac{d^3k'}{(2\pi)^32E_{k'}}$$
$$=\delta(x_1-x_B), \eqno(2.6)$$

in the TOPT-form.

\begin{center}
\section{Bare probe-vertex approximation}
\end{center}

As we know, emission or absorption of quanta with zero-momentum
may associate with the infrared (IR) divergence. However, the
singular terms provide the leading contributions to the DIS processes.
Therefore, a correct theory is IR-safe, where IR-divergences
are canceled, while the leading contributions are retained.
One way can to attain above two goals is to sum over cut diagrams
belonging to the same time-ordered uncut graph, since these graphs
have similar singular structure but may come up with opposite signs.

Deep inelastic scattering structure functions are the imaginary parts
of the amplitudes for the forward `Compton' scattering of the target with 
a probe. Using the time-ordered perturbative expansion of the statement 
of the unitarity of the S-matrix, one can prove that the structure 
functions are associated with the sum of cut diagrams.
These different cut graphs represent various possible
sub-partonic processes due to the unitarity of the perturbative
S-matrix [7]. Therefore, the sum of cut graphs is necessary not only
for infrared safety, but also for collecting the leading contributions
and restoring the unitarity.

The interesting and important question is, what are the minimum 
cut diagrams that must be summed for IR-safe calculation of an inclusive
DIS process at a given order (for example, order $\alpha_s^2/(RQ)^2 $
in this work) ?

	To  answer this question, let us consider a general inclusive
DIS process on target N. One can choose the cut diagrams according
to following program: G(N) stands for the time-ordered uncut diagram
of the target N without probe vertices. We sum over possible cut
diagrams of G(N):

$$\sum_\gamma G_\gamma(N)
=\{\sum_\gamma L_\gamma R_\gamma\}_I, \eqno(3.1)$$

where $G_\gamma$ is the cut diagram with cut line $\gamma$;
$L_\gamma$ and $R_\gamma$ are the sub-graphs on
the left and right of the cut line; the subscript ``I'' means that
we only consider following cut graphs which have the same observed quantities
(that is, the same structure of the intermediate state) and which keep the
original correlation among initial partons in $G(N)$.

We use the probe to observe the parton distributions inside the target
in DIS. Of course, we cannot control the probing positions.
In principle, in- and out-probe lines can be attached to the left- and 
right-hand of the cut line in all possible ways. 
Let $G^\beta_\gamma(probe+N)$ stand for the cut diagrams of the probe-target
system, where $\beta$ labels the probe-parton vertices. Thus, we
shall sum over 

$$\sum_\beta\sum_\gamma G_\gamma^{\beta}(probe+N). \eqno(3.2)$$

Obviously, the sum (3.2) is much larger than $\sum_\gamma G_\gamma(N)$.
Now we try to find some approximation in the sum (3.2). As we know
that the leading logarithmic approximation (LLA)
is a good approximation for IR-safeness in the DGLAP equation at
order-$\alpha_s$. In this approximation,
some of the renormalization effects are neglected in the physical
gauge and the probe-vertex retains the bare-vertex form as in (2.6).
In this case,

$$G^\beta_\gamma(probe+N)=G_\gamma(N)\delta(x_\gamma-x_B), \eqno(3.3)$$

Thus, the contributions from the nonlocal interactions of probe with
partons are neglected at the leading approximation. We need only
to sum part of cut graphs, in which the bare-probe vertex
$\delta(x_\gamma-x_B)$ connects with the cut line, that is,

$$\sum_\beta\sum_\gamma G^\beta_\gamma(probe+N)$$
$$=\sum_\gamma G_\gamma(N)\delta(x_\gamma-x_B)$$
$$=\sum_\gamma \{L_\gamma R_\gamma\}_I\delta(x_\gamma-x_B), \eqno(3.4)$$

in the DIS processes with parton recombination. 

We call (3.4) as the bare probe-vertex approximation.
We find that this approximation is also a satisfactory approximation 
in the DIS processes with parton recombination. In fact, our interest  
is that the modifications of the parton recombination to the DGLAP
equation, which has the probability explanation at the LLA approximation. 
We shall show that the bare probe-vertex approximation is necessary
for keeping the probability picture of the new evolution equation.

\begin{center}
\section{Rederivation of the DGLAP equation}
\end{center}

We know that several methods can be used to derive
the DGLAP evolution equations, however, the following new method
illustrates more clearly the simple relations among the cut diagrams in the
sum (3.4). For simplicity, we only consider the non-singlet case.
According to (3.4) we compute fig. 4a with figs. 4b and 4c.
Figures 4b and 4c seem to change the observed quantity $d\ln l_\perp^2$,
where $l_\perp$ is the transverse momentum of the final state partons, 
(see fig. 2a) whose momenta are parametrized as
$$l=\left [x_1p,\b{0},x_1p\right ],$$
$$k=\left [x_2p+\frac{l^2_{\perp}}{2x_2p},l_{\perp},x_2p \right ],$$
$$l'=\left [x_3p+\frac{l^2_{\perp}}{2x_3p},-l_{\perp},x_3p \right ].
\eqno(4.1)$$

However, we will see that the contribution of an intermediate state can be
replaced by $d^3l'$, which is from the contribution of the loop
and contributes $d\ln l_\perp^2$. Therefore, all the processes of fig. 4 
have the same intermediate state structure.

We proceed along the lines of ref.[8]. The change of the valence-quark-number 
density caused by gluon radiation can be written as (see fig. 4a)

$$dq(x_B)=\frac{1}{C_q}d\sigma(\gamma^* p\rightarrow k'X)$$
$$=\frac{E_l}{E_p}\vert M_{p\rightarrow lX}\vert^2[\frac{1}{E_p-E_l-E_X}]^2
[\frac{1}{2E_l}]^2
\prod_{X} \frac{d^3k_X}{(2\pi)^32E_X}$$
$$H(\gamma^*l\rightarrow \gamma^*l),\eqno(4.2)$$

where the cross section is factorized to the soft part
and the hard part $H(\gamma^*l\rightarrow \gamma^*l)$ according to the
factorization theory [9,10]. Using TOPT we obtain the hard part

$$H(\gamma^*l\rightarrow \gamma^*l)$$
$$=\frac{1}{C_q}\frac{1}{8E_kE_{\gamma}}\vert M_{{\gamma^*}k\rightarrow k'}
\vert^2 (2\pi)^4\delta^4(p_\gamma+k-k')\frac{d^3k'}{(2\pi)^3
2E_{k'}}$$
$$\frac{E_k}{E_l}\vert M_{l\rightarrow kl'}\vert^2[\frac{1}{E_l-E_k-E_{l'}}]^2
[\frac{1}{2E_k}]^2\frac{d^3l'}{(2\pi)^32E_{l'}}. \eqno(4.3)$$

Assuming that $l_{\perp}^2\sim Q^2$, we have
$$\frac {dq(x_B,Q^2)}{dlnQ^2}$$
$$=\int q(x_1,Q^2)P_2^{qq}(x_1,x_2,x_3)\delta(x_1-x_2-x_3)\delta(x_2-x_B)dx_1dx_2dx_3$$
$$=\int q(x_1,Q^2)P_2^{qq}(z)dz\delta(x_1z-x_B)dx_1$$
$$=\int q(x_1,Q^2)P_2^{qq}(\frac{x_B}{x_1})\frac{dx_1}{x_1},
\eqno(4.4)$$

where

$$q(x_1,Q^2)dx_1=\frac{E_l}{E_p}\vert M_{p\rightarrow lX}\vert^2
[\frac{1}{E_p-E_l-E_X}]^2[\frac{1}{2E_l}]^2\prod_{X} \frac{d^3k_X}
{(2\pi)^32E_X},  \eqno(4.5)$$

as same as eq.(2.5) and

$$\delta(x_2-x_B)=\frac{1}{C_q}\frac{1}{8E_kE_{\gamma}}
\vert M_{{\gamma^*}k\rightarrow k'}
\vert^2 (2\pi)^4\delta^4(p_{\gamma}+k-k')\frac{d^3k'}{(2\pi)^3
2E_{k'}}.\eqno(4.6)  $$

In eq.(4.4) we
inserted
$$\int\delta(x_1-x_2-x_3)dx_2=1. \eqno(4.7)$$

We define

$$P_2^{qq}(x_1,x_2,x_3)dx_3\frac{dl^2_{\perp}}{l^2_{\perp}}=
\frac{E_k}{E_l}\vert M_{l\rightarrow kl'}\vert^2[\frac{1}{E_l-E_k-E_{l'}}]^2
[\frac{1}{2E_k}]^2\frac{d^3l'}{(2\pi)^32E_{l'}},\eqno(4.8)$$

as the parton splitting function for the non-singlet part. 

Now let us consider figs. 4b and 4c. As for the
the real diagram, the contributions of fig. 4b to the change
of the valence-quark-number density are

$$dq(x_B)=\frac{1}{2}\frac{1}{C_q}d\sigma(\gamma^* p\rightarrow k'X)$$
$$=\frac{1}{2}\frac{E_l}{E_p}\vert M_{p\rightarrow lX}\vert ^2
[\frac{1}{E_p-E_l-E_X}]^2[\frac{1}{2E_l}]^2
\prod_X\frac{d^3k_X}{(2\pi)^32E_X}$$
$$H(\gamma^*l\rightarrow \gamma^*l), \eqno(4.9)$$

where the factor of $\frac{1}{2}$ was explained as the effect of 
the renormalization in ref. [11]:
The virtual part of fig.4b corresponds to the renormalization  
of a parton propagator. Only half of the probe-vertex connects with this 
parton line. This is equivalent to multiplying the virtual process 
by an extra factor of $\frac{1}{2}$. 

	The hard part in eq.(4.9) is

$$H(\gamma^*l\rightarrow \gamma^*l)$$
$$=\frac{1}{C_q}\frac{1}{8E_lE_{\gamma}}\vert M_{{\gamma^*}l\rightarrow k'}
\vert^2 (2\pi)^4\delta^4(p_{\gamma}+l-k')\frac{d^3k'}{(2\pi)^32E_{k'}}$$
$$\frac{d^3l'}{(2\pi)^3}M_{l\rightarrow kl'}\frac{1}{2E_k}\frac{1}{2E_{l'}}
\frac{1}{E_l-E_k-E_{l'}}M_{kl'\rightarrow l}\frac{1}{E_k+E_{l'}-E_l}
\frac{1}{2E_l}$$
$$=-\frac{1}{C_q}\frac{1}{8E_lE_{\gamma}}\vert M_{\gamma^*l\rightarrow k'}
\vert^2 (2\pi)^4\delta^4(p_{\gamma}+l-k')\frac{d^3k'}{(2\pi)^32E_{k'}}$$
$$\frac{E_k}{E_l}\vert M_{l\rightarrow kl'}\vert^2[\frac{1}
{E_l-E_k-E_{l'}}]^2
[\frac{1}{2E_k}]^2\frac{d^3l'}{(2\pi)^32E_{l'}}. \eqno(4.10)$$

Therefore we have

$$\frac {dq(x_B,Q^2)}{dlnQ^2}$$
$$=-\int \frac{1}{2}q(x_1,Q^2)P_2^{qq}(x_1,x_2,x_3)\delta(x_1-x_2-x_3)
\delta(x_1-x_B)dx_1dx_2dx_3$$
$$=-\int \frac{1}{2}q(x_1,Q^2)P_2^{qq}(z)dz\delta(x_1-x_B)dx_1$$
$$=-\frac{1}{2}q(x_B,Q^2)\int P_2^{qq}(z)dz,
\eqno(4.11)$$

in which we have used eqs.(2.5),(2.6) and (4.8).

The contributions of fig. 4c is same as one of fig. 4b. Thus, the
total contributions of the real- and virtual-diagrams are

$$\frac {dq(x_B,Q^2)}{dlnQ^2}=\int q(x_1,Q^2)P_2^{qq}(\frac{x_B}{x_1})
\frac{dx_1}{x_1}-q(x_B,Q^2)\int P_2^{qq}(z)dz, \eqno(4.12)$$

in which

$$P_2^{qq}(z)=\frac{\alpha_s}{2\pi}C_2(R)\frac{1+z^2}{1-z}. \eqno(4.13)$$

The two terms of the right-hand side in eq.(4.12) have a simple 
interpretation: the positive contribution
arises from the splitting of higher momentum quarks, while the
negative contribution results in the loss of the number of quarks due to
its gluon radiation.

The result (4.12) is the same as the probability form of the DGLAP equation
for the nonsinglet part in ref. [11]. However, the new derivation 
clearly shows the following interesting properties in the inclusive 
DIS processes: The contributions of the cut diagrams, which belong to 
a same time-ordered uncut graph in the sum (3.4),
have an identical integral kernel (it is the parton
splitting function in eq.(4.12)). This is a reason we use the TOPT
form to perform our calculations. We shall examine this 
connection further in the parton recombination processes.

\begin{center}
\section{Leading recombination approximation}
\end{center}

So far we considered processes contributing to the usual DGLAP equation.
We go on now to include parton recombination processes.
The recombination processes contributing at leading order come 
from the terms, 
$\vert M^{(2-2)}_{p\gamma^*\rightarrow k'l'X}\vert ^2$,
$2M^{(2-1)}_{p\gamma^*\rightarrow
k'l'X}[M^{(2-3)}_{p\gamma^*\rightarrow k'l'X}]^*$, and the cut
diagrams according to (3.4).

	In this section we regard the partons as scalar particles 
(i.e., the $\phi^3$ model). The results can easily be generalized to 
the case of QCD partons and will be done later in section 7. 
We consider the process of fig. 5, where figs. 5c--f are virtual.
The contribution of the real diagram fig. 5a is

$$d\phi(x_B)=\frac{1}{C_\phi}d\sigma(\gamma^* p\rightarrow k'X)$$
$$=\frac{\sqrt{E_{p_1}+E_{p_2}}\sqrt{E_{p'_1}+E_{p'_2}}}{E_p}
M_{p\rightarrow p_1p_2X}[M_{p\rightarrow p'_1p'_2X}]^*$$
$$(\frac{1}{E_P-E_{p_1}-E_{p_2}-E_X})^2
\frac{1}{2E_{p_1}}\frac{1}{2E_{p_2}}\frac{1}{2E_{p'_1}}\frac{1}{2E_{p'_2}}
\prod_{X} \frac{d^3k_X}{(2\pi)^32E_X}$$
$$H(\gamma^*p_1p_2\rightarrow \gamma^*p'_1p'_2),
\eqno(5.1)$$

where $C_\phi$ is defined by

$$d\sigma(\gamma^*k\rightarrow k')=C_\phi\delta(x_1-x_B), \eqno (5.2)$$

for the scalar parton; the hard part is given by,

$$H(\gamma^*p_1p_2\rightarrow \gamma^*p'_1p'_2)$$
$$=(\frac{1}{R})^2\frac{1}{C_\phi}\frac{1}{8E_kE_{\gamma}}
\vert M_{{\gamma^*}k\rightarrow k'}\vert^2
(2\pi)^4\delta^4(p_\gamma+k-k')\frac{d^3k'}{(2\pi)^32E_{k'}}$$
$$\frac{E_k}{\sqrt{E_{p_1}+E_{p_2}}\sqrt{E_{p'_1}+E_{p'_2}}}
M_{p_1p_2\rightarrow kl'}[M_{p'_1p'_2\rightarrow kl'}]^*
(\frac{1}{E_{p_1}+E_{p_2}-E_k-E_{l'}})^2$$
$$(\frac{1}{2E_k})^2\frac{d^3l'}{(2\pi)^32E_{l'}}.
\eqno(5.3)$$

We define the parton correlation function (PCF) $f(x_1,x_2;x'_1,x'_2)$
as [12],

$$f(x_1,x_2;x'_1,x'_2)\delta(x_1+x_2-x'_1-x'_2)dx_1dx_1'dx_2dx_2'$$
$$=\frac{\sqrt{E_{p_1}+E_{p_2}}\sqrt{E_{p'_1}+E_{p'_2}}}{E_p}
M_{p\rightarrow p_1p_2X}[M_{p\rightarrow p'_1p'_2X}]^*
(\frac{1}{E_p-E_{p_1}-E_{p_2}-E_{X}})^2$$
$$\frac{1}{2E_{p_1}}\frac{1}{2E_{p_2}}\frac{1}{2E_{p'_1}}\frac{1}{2E_{p'_2}}
\prod_{X} \frac{d^3k_X}{(2\pi)^32E_X},
\eqno(5.4)$$

while the parton recombination function $P_4^{(2-2)}$ is defined by

$$P_4^{(2-2)}(x_1,x_2,x'_1,x'_2,x_3,x_4)dx_4
\frac{dl^2_{\perp}}{l^4_{\perp}}$$
$$=\frac{E_k}{\sqrt{E_{p_1}+E_{p_2}}\sqrt{E_{p'_1}+E_{p'_2}}}
M_{p_1p_2\rightarrow kl'}[M_{p'_1p'_2\rightarrow kl'}]^*
(\frac{1}{E_{p_1}+E_{p_2}-E_k-E_{l'}})^2$$
$$(\frac{1}{2E_k})^2\frac{d^3l'}{(2\pi)^32E_{l'}}.
\eqno(5.5)$$

We shall discuss the PCF and the parton recombination function in
sections 6 and 7, respectively. Therefore, we have

$$\frac {d\phi(x_B,Q^2)}{d\ln Q^2}$$
$$=(\frac{1}{RQ})^2\int f(x_1,x_2,x'_1,x'_2,Q^2)
\delta(x_1+x_2-x'_1-x'_2)P_4^{(2-2)}(x_1,x_2,x'_1,x'_2,x_3,x_4)$$
$$\delta(x_3-x_B)\delta(x_1+x_2-x_3-x_4)dx_1dx_2dx'_1dx'_2dx_3dx_4,
\eqno(5.6)$$

We shall discuss the PCF and the parton recombination function in
sections 6 and 7, respectively. Therefore, we have
$$\frac {dq(x_B,Q^2)}{d\ln Q^2}$$
$$=(\frac{1}{RQ})^2\int f(x_1,x_2,x'_1,x'_2,Q^2)
\delta(x_1+x_2-x'_1-x'_2)P_4^{(2-2)}(x_1,x_2,x'_1,x'_2,x_3,x_4)$$
$$\delta(x_3-x_B)\delta(x_1+x_2-x_3-x_4)dx_1dx_2dx'_1dx'_2dx_3dx_4,
\eqno(5.5)$$

where we have inserted a factor $1 = \int \delta(x_1+x_2-x_3-x_4)dx_3$.
This is the evolution equation from fig. 5a.

Similarly, using the factorization in DIS [9,10], the contribution 
of fig. 5c (virtual diagram) is

$$d\phi(x_B)=\frac{1}{2}\frac{1}{C_\phi}d\sigma(\gamma^* p\rightarrow k'X)$$
$$=\frac{1}{2}\frac{\sqrt{E_{p_1}+E_{p_2}}
\sqrt{E_{p'_1}+E_{p'_2}}}{E_p}
M_{p\rightarrow p_1X}[M_{p\rightarrow p'_1p'_2p_2X}]^*$$
$$\frac{1}{E_p-E_{p_1}-E_X}(\frac{1}{E_p-E_{p'_1}-E_{p'_2}-E_{p_2}-
E_X})^*
\frac{1}{2E_{p_1}}\frac{1}{2E_{p_2}}\frac{1}{2E_{p'_1}}\frac{1}{2E_{p'_2}}
\prod_{X} \frac{d^3k_X}{(2\pi)^32E_X}$$
$$H(\gamma^*p_1p_2\rightarrow \gamma^*p'_1p'_2).
\eqno(5.7)$$

Now the PCF is defined by

$$f(x_1;x_2,x'_1,x'_2)dx_1dx_1'dx_2dx_2'$$
$$=\frac{\sqrt{E_{p_1}+E_{p_2}}
\sqrt{E_{p'_1}+E_{p'_2}}}{E_p}
M_{p\rightarrow p_1X}[M_{p\rightarrow p'_1p'_2p_2X}]^*$$
$$\frac{1}{E_p-E_{p_1}-E_X}(\frac{1}{E_p-E_{p'_1}-E_{p'_2}-E_{p_2}-
E_X})^*$$
$$\frac{1}{2E_{p_1}}\frac{1}{2E_{p_2}}\frac{1}{2E_{p'_1}}\frac{1}{2E_{p'_2}}
\prod_{X} \frac{d^3k_X}{(2\pi)^32E_X}.
\eqno(5.8)$$

However, we have the condition,

$$f(x_1;x_2,x'_1,x'_2)=f(x_1,x_2;x'_1,x'_2),
\eqno(5.9)$$

since the PCFs with cuts at different places are the same on the light-cone
(fig.6) [13,14].

The hard part is given by

$$H(\gamma^*p_1p_2\rightarrow \gamma^*p'_1p'_2)$$
$$=(\frac{1}{R})^2\frac{1}{C_\phi}\frac{1}{8E_{\gamma}}
\vert M_{{\gamma^*}p_1\rightarrow k'}\vert^2
(2\pi)^4\delta^4(p_\gamma+p_1-k')\frac{d^3k'}{(2\pi)^32E_{k'}}$$
$$\frac{1}{\sqrt{E_{p_1}+E_{p_2}}\sqrt{E_{p'_1}+E_{p'_2}}}
\frac{1}{2E_{p_1}}\frac{d^3l'}{(2\pi)^3}M_{p_1p_2\rightarrow kl'}
[M_{p'_1p'_2\rightarrow kl'}]^* $$
$$(\frac{1}{E_{p_1}+E_{p_2}-E_k-E_{l'}})^*
(\frac{1}{E_k+E_{l'}-E_{p'_1}-E_{p'_2}})^*
(\frac{1}{2E_k})^*(\frac{1}{2E_{l'}})^*$$
$$=-(\frac{1}{R})^2\frac{1}{C_\phi}\frac{1}{8E_{p_1}E_{\gamma}}
\vert M_{{\gamma^*}p_1\rightarrow k'}\vert^2
(2\pi)^4\delta^4(p_\gamma+p_1-k')\frac{d^3k'}{(2\pi)^32E_{k'}}$$
$$\frac{E_k}{\sqrt{E_{p_1}+E_{p_2}}\sqrt{E_{p'_1}+E_{p'_2}}}
M_{p_1p_2\rightarrow kl'}[M_{p'_1p'_2\rightarrow kl'}]^*
(\frac{1}{E_{p_1}+E_{p_2}-E_k-E_{l'}})^2$$
$$(\frac{1}{2E_k})^2\frac{d^3l'}{(2\pi)^32E_{l'}},
\eqno(5.10)$$

where the factor of $\frac{1}{2}$ in (5.7) is needed for the cancellation 
of IR-divergences and momentum conservation as we shall discuss shortly.
One can re-understand this factor as follows: 
only half of the probe-vertex connects with the partonic matrix 
in figs. 5c-f as well as in figs. 4b-c,
and the square root of the parton density accepts the
contributions of the partonic processes through a parton line. That is,

$$\frac{\sqrt {q}d\sqrt {q}}{d\ln Q^2}=\frac{1}{2}\frac {dq}{d\ln Q^2}.
\eqno(5.11) $$

	Since, for the given initial partons we have,

$$\sum_{s,t,u}M_{p_1p_2\rightarrow kl'}[M_{p'_1p'_2\rightarrow kl'}]^* >0,
\eqno(5.12)$$

we can conclude that the negative sign in eq. (5.10) arises from

$$\frac{1}{E_{p_1}+E_{p_2}-E_k-E_{l'}}\frac{1}{E_k+E_{l'}-E_{p'_1}-
E_{p'_2}}
=-(\frac{1}{E_{p_1}+E_{p_2}-E_k-E_{l'}})^2.
\eqno(5.13)$$

In consequence, we have

$$\frac {d\phi(x_B,Q^2)}{d\ln Q^2}
=-\frac{1}{2}(\frac{1}{RQ})^2
\int f(x_1,x_2;x'_1,x'_2,Q^2)\delta(x_1+x_2-x'_1-x'_2)$$
$$P_4^{(2-2)}(x_1,x_2,x'_1,x'_2,x_3,x_4)$$
$$\delta(x_1-x_B)\delta(x_1+x_2-x_3-x_4)dx_1dx_2dx'_1dx'_2dx_3dx_4.
\eqno(5.14)$$

Comparing eqs. (5.14) with (5.6), we see that the same   
recombination function appears in both cases.  
We can calculate the contributions of figs. 5a--f 
in TOPT using the same method, and finally obtain
the hard contribution as 

$$\frac {d\phi(x_B,Q^2)}{d\ln Q^2}
=(\frac{1}{RQ})^2
\int f(x_1,x_2;x'_1,x'_2,Q^2)P_4^{(2-2)}(x_1,x_2,x'_1,x'_2, x_3,x_4)$$
$$\delta(x_1+x_2-x'_1-x'_2)[\delta(x_3-x_B)+\delta(x_4-x_B)
-\frac{1}{2}\delta(x_1-x_B)$$
$$-\frac{1}{2}\delta(x'_1-x_B)-\frac{1}{2}\delta(x_2-x_B)-
\frac{1}{2}\delta(x'_2-x_B)]$$
$$\delta(\frac{1}{2}x_1+\frac{1}{2}x'_1+\frac{1}{2}x_2+\frac{1}{2}x'_2-x_3-x_4)
dx_1dx_2dx'_1dx'_2dx_3dx_4.
\eqno(5.15)$$

     Obviously, eq. (5.14) contains the momentum conservation condition:

$$\frac {d\displaystyle \int_{0}^{1}x_B\phi(x_B,Q^2)dx_B}{d\ln Q^2}=0.
\eqno(5.16)$$

This completes the discussion of the first type of diagrams.
As the next step, we discuss the interference terms,
$M^{(2-1)}_{p\gamma^*\rightarrow k'l'X}[M^{(2-3)}_{p\gamma^*\rightarrow k'l'X}]^*$,
shown in fig. 7. Proceeding similarly, we obtain the 
contributions from the interference processes in fig. 7 to be

$$\frac {d\phi(x_B,Q^2)}{d\ln Q^2}
=2P_{inter}(\frac{1}{RQ})^2
\int f(x_1;x_2,x'_1,x'_2,Q^2)P_4^{(1-3)}(x_1,x_2,x'_1,x'_2, x_3,x_4)$$
$$\delta(x_1-x_2-x'_1-x'_2)[\delta(x_3-x_B)+\delta(x_4-x_B)
-\frac{1}{2}\delta(x_1-x_B)$$
$$-\frac{1}{2}\delta(x'_1-x_B)-\frac{1}{2}\delta(x_2-x_B)-
\frac{1}{2}\delta(x'_2-x_B)]$$
$$\delta(x_1-x_3-x_4)
dx_1dx_2dx'_1dx'_2dx_3dx_4,
\eqno(5.17)$$

where

$$f(x_1;x_2,x'_1,x'_2,Q^2)
=\frac{\sqrt{E_{p_1}+E_{p_2}}
\sqrt{E_{p'_1}+E_{p'_2}}}{E_p}
M_{p\rightarrow p_1X}[M_{p\rightarrow p'_1p'_2p_2X}]^*$$
$$\frac{1}{E_p-E_{p_1}-E_{X'}}(\frac{1}{E_p-E_{p'_1}-E_{p'_2}-E_{p_2}-
E_X})^*
\frac{1}{2E_{p_1}}\frac{1}{2E_{p_2}}\frac{1}{2E_{p'_1}}\frac{1}{2E_{p'_2}}
\prod_{X} \frac{d^3k_X}{(2\pi)^32E_X}$$
$$H(\gamma^*p_1p_2\rightarrow \gamma^*p'_1p'_2).
\eqno(5.18)$$

and

$$P_4^{(1-3)}(x_1,x_2,x'_1,x'_2,x_3,x_4)dx_4
\frac{dl^2_{\perp}}{l^4_{\perp}}$$
$$=\frac{E_k}{\sqrt{E_{p_1}+E_{p_2}}\sqrt{E_{p'_1}+E_{p'_2}}}
M_{p_1\rightarrow kl'}[M_{p'_1p'_2p_2\rightarrow kl'}]^*$$
$$\frac{1}{E_k+E_l'-E_{p_1}}\frac{1}{E_k+E_{l'}-E_{p'_1}-E_{p'_2}-E_{p_2}}
(\frac{1}{2E_k})^2\frac{d^3l'}{(2\pi)^32E_{l'}}.
\eqno(5.19)$$

In eq. (5.17), $P_{inter}=0$ or 1 implies that the interference processes
are inhibited or exhibited, respectively.
Now an interesting observation is that $P^{(2-2)}_4$ and $P^{(1-3)}_4$
are really similar except for a simple coefficient and the different 
variable range. In fact, from fig. 8 we find

$$P^{(2-2)}_4=-\xi P^{(1-3)}_4,
\eqno(5.20)$$

where $\xi$ is defined below and the negative sign occurs because
$({l_{\perp}^2}/(2x_lp)- {l_{\perp}^2}/(2x_{l'}p))$ changes its
sign from $x_l<x_{l'}$ to $x_l>x_{l'}$ in fig. 8.
The contributions of the vertices A and B have same form, since the 
momenta of the partons $a$ and $b$ are 
$\left [x_ap,\b{0},x_ap\right ]$ and 
$\left [x_bp,\b{0},x_bp\right ]$, respectively.
The factor $\xi$ arises from the following
symmetry: if both the final partons are gluons or
quarks (we do not distinguish quarks and antiquarks in this work), the
corresponding virtual diagrams in figs. 7c-f are symmetric under the exchange
of these two partons. However, this symmetry
will be lost if we use $P^{(2-2)}_4$ to replace $P^{(1-3)}_4$. In this case,
$\xi=1/2$, otherwise, $\xi=1$.

	It seems that there are
different energy deficits in going from $P^{(1-3)}_4$ to $P^{(2-2)}_4$
in the ATOPT: ${1}/(E_k+E_{l'}-E_{p_1})$ in $P^{(1-3)}_4$ and
${1}/(E_k+E_{l'}-E_{p_1}-E_{p_2})$ in $P^{(2-2)}_4$;
${1}/(E_k+E_{l'}-E_{p'_1}-E_{p'_2}-E_{p_2})$ in $P^{(1-3)}_4$ and
${1}/(E_k+E_{l'}-E_{p'_1}-E_{p'_2})$ in $P^{(2-2)}_4$. However, they
are really the same factor, arising from the term,
$(l^2_{\perp}/(2x_3p)+l^2_{\perp}/(2x_4p))^{-1}$.
	
	Therefore, one can replace $P^{(1-3)}_4$ by $P^{(2-2)}_4$
and reconstruct eq. (5.17), where the terms
$f(x_1;x'_1,x'_2,x_2,Q^2)$ and
$\delta(x_1-x'_1-x'_2-x_2)$ should be replaced by
$f(x_1,x_2;x'_1,x'_2,Q^2)$ and
$\delta(x_1+x_2-x'_1-x'_2)$, respectively, since $x_i$ are the scaling
variables.

	The final results from fig. 7 are

$$\frac {d\phi(x_B,Q^2)}{d\ln Q^2}$$
$$=2(\frac{1}{RQ})^2P_{inter}
\int f(x_1,x_2;x'_1,x'_2,Q^2)P_4^{(2-2)}(x_1,x_2,x'_1,x'_2, x_3,x_4)$$
$$\delta(x_1+x_2-x'_1-x'_2)[-\delta(x_3-x_B)-\delta(x_4-x_B)
+\frac{1}{2}\delta(x_1-x_B)$$
$$+\frac{1}{2}\delta(x'_1-x_B)+\frac{1}{2}\delta(x_2-x_B)+
\frac{1}{2}\delta(x'_2-x_B)]$$
$$\delta(\frac{1}{2}x_1+\frac{1}{2}x'_1+\frac{1}{2}x_2+\frac{1}{2}x'_2-x_3-x_4)
dx_1dx_2dx'_1dx'_2dx_3dx_4.
\eqno(5.21)$$

Obviously, the momentum conservation condition is also satisfied in eq. (5.21),

$$\frac {d\displaystyle \int_{0}^{1}x_B\phi(x_B,Q^2)dx_B}
{d\ln Q^2}=0.
\eqno(5.22)$$

\begin{center}
\section{Parton correlation functions}
\end{center}

In general, the parton density is a concept that is only defined
at the twist-2 level; it can be expressed in terms of the product of
the initial and final hadronic wave functions with the same parton 
configuration. The parton correlation function
$f(x_1,x_2;x'_1,x'_2)$ is a generalization of the parton density
beyond the leading twist. It has not yet been experimentally observed.
In this section, therefore, we shall try to construct the
connection between the parton correlation function and the parton density.

For example, consider the correlation function for the case when 
$x_1=x'_1$ and $x_2=x'_2$ in eq.(5.8); this is given by

$$f(x_1,x_2;x_1,x_2)$$
$$=\frac{E_{p_1}E_{p_2}}{E_p}
\vert M_{p\rightarrow p_1p_2X}\vert^2
(\frac{1}{E_{p_1}+E_{p_2}+E_X-E_p})^2
(\frac{1}{2E_{p_1}})^2(\frac{1}{2E_{p_2}})^2
\prod_{X} \frac{d^3k_X}{(2\pi)^32E_X}$$
$$=\rho(x_1,x_2),
\eqno(6.1)$$

and is the number density of two partons, i.e., the probability of
simultaneously finding two partons carrying $x_1$ and $x_2$ fractions of
the proton momentum respectively. In the quantum mechanics
approximation, we can use wave functions to represent $\rho (x_1,x_2)$ as

$$\rho(x_1,x_2)=\psi(x_1,x_2)\psi^*(x_1,x_2),
\eqno(6.2)$$

where $\psi(x_1,x_2)$ is the wave function of two partons in the proton.
Similarly, we express the parton correlation function as the
product of the initial and final hadron wave functions with different
parton momentum, that is

$$f(x_1,x_2;x'_1, x'_2)
=\psi(x_1,x_2)\psi^*(x'_1,x'_2).
\eqno(6.3)$$

We define

$$\overline{f}(x_1,x_2;x'_1, x'_2)
\equiv \psi (x'_1, x'_2)\psi^*(x_1,x_2)
=\kappa^2 f(x_1,x_2;x'_1, x'_2). \eqno(6.4)$$

Therefore, we have

$$f(x_1,x_2;x'_1, x'_2)=\kappa^{-1}\sqrt{f(x_1,x_2;x'_1, x'_2)
\overline{f}(x_1,x_2;x'_1, x'_2)}$$
$$=\kappa^{-1}\sqrt{\psi(x_1,x_2)\psi^*(x_1,x_2)\psi(x'_1,x'_2)
\psi^*(x'_1, x'_2)}$$
$$=\kappa^{-1}\sqrt{\rho(x_1,x_2)\rho(x'_1,x'_2)}.
\eqno(6.5)$$
In general, the two-parton number density can be approximated by

$$\rho(x_1,x_2)=q_a(x_1)q_b(x_2)R^2_{ab}(x_1,x_2),
\eqno(6.6)$$

where $R^2_{ab}(x_1,x_2)$ is the momentum correlation of
two initial partons;
$q_a(x_1)$ and $q_b(x_2)$ are the parton number densities.

In order to estimate the value of $\kappa$ in eq. (6.5), we consider
the process shown in fig. 9. The time reversal invariance requires that
$$d\sigma(\gamma^*p\rightarrow \gamma^*p)=d\sigma(\gamma^*p\leftarrow
\gamma^*p),
\eqno(6.7)$$

or

$$\psi(x_1,x_2)AB^*\psi^*(x'_1,x'_2)=\psi(x'_1,x'_2)BA^*\psi^*(x_1,x_2),
\eqno(6.8)$$

where A and B are the contributions of the hard parts in fig. 9.
Therefore,

$$\kappa=\sqrt{\frac {\overline{f}(x_1,x_2;x'_1,x'_2)}
{f(x_1,x_2;x'_1,x'_2)}}$$
$$=\sqrt{\frac{AB^*}{BA^*}}.
\eqno(6.9)$$

Since $\kappa$ is expressible in terms of hard parts, eq. (6.9) indicates
that $\kappa$ is calculable within PQCD.

\begin{center}
\section{New evolution equations}
\end{center}

We now apply the method, used to describe scalar partons in Section 5, to
the realistic case of partons (quarks and gluons) interacting within QCD.
In consequence, we have following new evolution equations with twist-4
for $GG\rightarrow q\overline{q}$ and $GG\rightarrow GG$ respectively:

\paragraph{(a) $GG\rightarrow q\overline{q}$.}
The contribution to the evolution equation for gluons is,

$$\frac{dG(x_B,Q^2)}{d\ln Q^2}$$
$$=(\frac{1}{RQ})^2\int \sqrt{G(x_1,Q^2)G(x_2,Q^2)G(x_1+
\Delta,Q^2)G(x_2-\Delta,Q^2)}$$
$$R_{GG}(x_1, x_2)R_{GG}(x_1+\Delta, x_2-\Delta)$$
$$\sum_i\kappa^{-1}_iP_{GG\rightarrow q\overline{q}}^i(x_1,x_2,x_3,x_4,\Delta)\delta(x_1+x_2-x_3-x_4)$$
$$[-\frac{1}{2}\delta(x_1-x_B)-\frac{1}{2}\delta(x_1+\Delta-x_B)
-\frac{1}{2}\delta(x_2-x_B)-\frac{1}{2}\delta(x_2-\Delta-
x_B)]dx_1dx_2dx_3dx_4d\Delta$$
$$+2(\frac{1}{RQ})^2P_{inter.}\int
\sqrt{G(x_1,Q^2)G(x_2,Q^2)G(x_1+\Delta,Q^2)G(x_2-\Delta
,Q^2)}$$
$$R_{GG}(x_1, x_2)R_{GG}(x_1+\Delta, x_2-\Delta)$$
$$\sum_i\kappa^{-1}_iP_{GG\rightarrow q\overline{q}}^i(x_1,x_2,x_3,x_4,\Delta)\delta(x_1+x_2-x_3-x_4)$$
$$\frac{1}{2}[\frac{1}{2}\delta(x_1-x_B)+\frac{1}{2}\delta(x_1+\Delta-x_B)
+\frac{1}{2}\delta(x_2-x_B)+\frac{1}{2}\delta(x_2-\Delta-
x_B)]dx_1dx_2dx_3dx_4d\Delta, 
\eqno(7.1)$$
 
where the factor $\frac{1}{2}$ in the last factor arises from symmetry
considerations, just as in eq. (5.20). However, this
symmetry will be broken due to the cut in real diagrams in the
corresponding equation for quarks:

$$\frac{dq(x_B,Q^2)}{d\ln Q^2}$$
$$=(\frac{1}{RQ})^2\int
\sqrt{G(x_1,Q^2)G(x_2,Q^2)G(x_1+\Delta,Q^2)G(x_2-\Delta,Q^2)}$$
$$R_{GG}(x_1, x_2)R_{GG}(x_1+\Delta, x_2-\Delta)$$
$$\sum_i\kappa^{-1}_iP_{GG\rightarrow q\overline{q}}^i(x_1,x_2,x_3,x_4,\Delta)\delta(x_1+x_2-x_3-x_4)
[\delta(x_3-x_B)+\delta(x_4-x_B)]dx_1dx_2dx_3dx_4d\Delta$$
$$+2(\frac{1}{RQ})^2P_{inter}
\int \sqrt{G(x_1,Q^2)G(x_2,Q^2)G(x_1+\Delta,Q^2)G(x_2-\Delta,Q^2)}$$
$$R_{GG}(x_1, x_2)R_{GG}(x_1+\Delta, x_2-\Delta)$$
$$\sum_i\kappa^{-1}_iP_{GG\rightarrow q\overline{q}}^i(x_1,x_2,x_3,x_4,\Delta)\delta(x_1+x_2-x_3-x_4)$$
$$[-\delta(x_3-x_B)-\delta(x_4-x_B)]dx_1dx_2dx_3dx_4d\Delta.
\eqno(7.2)$$

\paragraph{(b) $GG\rightarrow GG$.}
The contribution to the evolution equation for gluons is,

$$\frac{dG(x_B,Q^2)}{d\ln Q^2}$$
$$=(\frac{1}{RQ})^2
\int \sqrt{G(x_1,Q^2)G(x_2,Q^2)G(x_1+\Delta,Q^2)G(x_2-\Delta,Q^2)}$$
$$R_{GG}(x_1, x_2)R_{GG}(x_1+\Delta, x_2-\Delta)$$
$$\sum_i\kappa^{-1}_iP_{GG\rightarrow GG}^i(x_1,x_2,x_3,x_4,\Delta)\delta(x_1+x_2-x_3-x_4)$$
$$[\delta(x_3-x_B)+\delta(x_4-x_B)-\frac{1}{2}\delta(x_1-x_B)-
\frac{1}{2}\delta(x_1+\Delta-
x_B)-\frac{1}{2}\delta(x_2-x_B)-\frac{1}{2}\delta(x_2-\Delta-x_B)]$$
$$dx_1dx_2dx_3dx_4d\Delta$$
$$+2(\frac{1}{RQ})^2P_{inter}
\int \sqrt{G(x_1,Q^2)G(x_2,Q^2)G(x_1+\Delta,Q^2)G(x_2-\Delta ,Q^2)}$$
$$R_{GG}(x_1, x_2)R_{GG}(x_1+\Delta, x_2-\Delta)$$
$$\sum_i\kappa^{-1}_iP_{GG\rightarrow GG}^i(x_1,x_2,x_3,x_4,\Delta)\delta(x_1+x_2-x_3-x_4)$$
$$[-\delta(x_3-x_B)-\delta(x_4-x_B)+\frac{1}{4}\delta(x_1-x_B)+
\frac{1}{4}\delta(x_1+\Delta+
x_B)+\frac{1}{4}\delta(x_2-x_B)+\frac{1}{4}\delta(x_2-\Delta-x_B)]$$
$$dx_1dx_2dx_3dx_4d\Delta. 
\eqno(7.3)$$

	We discuss the case when $P_{inter} = 1$ and $0$ separately.

\paragraph{A: $P_{inter}=1$.} We can take $P_{inter}=1$ if there is no
reason to forbid three-parton recombination in the
interference terms in nucleon. Thus, we have

$$\frac{dG(x_B,Q^2)}{d\ln Q^2}$$
$$=(\frac{1}{RQ})^2
\int_{(x_1+x_2)\geq x_B} \sqrt{G(x_1,Q^2)G(x_2,Q^2)G(x_1+\Delta,Q^2)G(x_2-\Delta,Q^2)}$$
$$R_{GG}(x_1, x_2)R_{GG}(x_1+\Delta, x_2-\Delta)$$
$$\sum_i\kappa^{-1}_iP_{GG\rightarrow GG}^i(x_1,x_2,x_3,x_4,\Delta)
\delta(x_1+x_2-x_3-x_4)$$
$$[\delta(x_3-x_B)+\delta(x_4-x_B)]dx_1dx_2dx_3dx_4d\Delta$$
$$-2(\frac{1}{RQ})^2
\int_{x_1\geq x_B} \sqrt{G(x_1,Q^2)G(x_2,Q^2)G(x_1+\Delta,Q^2)G(x_2-\Delta ,Q^2)}$$
$$R_{GG}(x_1, x_2)R_{GG}(x_1+\Delta, x_2-\Delta)$$
$$\sum_i\kappa^{-1}_iP_{GG\rightarrow GG}^i(x_1,x_2,x_3,x_4,\Delta)\delta(x_1+x_2-x_3-x_4)$$
$$[\delta(x_3-x_B)+\delta(x_4-x_B)]dx_1dx_2dx_3dx_4d\Delta. 
\eqno(7.4)$$

and

$$\frac{dq(x_B,Q^2)}{d\ln Q^2}$$
$$=(\frac{1}{RQ})^2
\int_{(x_1+x_2)\geq x_B} \sqrt{G(x_1,Q^2)G(x_2,Q^2)G(x_1+\Delta,Q^2)G(x_2-\Delta,Q^2)}$$
$$R_{GG}(x_1, x_2)R_{GG}(x_1+\Delta, x_2-\Delta)$$
$$\sum_i\kappa^{-1}_iP_{GG\rightarrow q\overline{q}}^i(x_1,x_2,x_3,x_4,\Delta)
\delta(x_1+x_2-x_3-x_4)$$
$$[\delta(x_3-x_B)+\delta(x_4-x_B)]dx_1dx_2dx_3dx_4d\Delta$$
$$-2(\frac{1}{RQ})^2
\int_{x_1\geq x_B} \sqrt{G(x_1,Q^2)G(x_2,Q^2)G(x_1+\Delta,Q^2)G(x_2-\Delta ,Q^2)}$$
$$R_{GG}(x_1, x_2)R_{GG}(x_1+\Delta, x_2-\Delta)$$
$$\sum_i\kappa^{-1}_iP_{GG\rightarrow q\overline{q}}^i(x_1,x_2,x_3,x_4,\Delta)\delta(x_1+x_2-x_3-x_4)$$
$$[\delta(x_3-x_B)+\delta(x_4-x_B)]dx_1dx_2dx_3dx_4d\Delta. 
\eqno(7.5)$$

	Equations (7.4) and (7.5) predict that the shadowing effect 
in quark distributions is stronger than that in the gluon distribution, 
since there are two shadowing sources for quarks but only one shadowing 
source for gluons.

\paragraph{B: $P_{inter}=0$.} This means that the interference
terms are forbidden. An example of such a case is the radiation recombination
in a nucleus. We consider the recombination of partons which originate
from different nucleons in a nucleus. A single parton can not escape
from the confinement region of a nucleon, unless it forms a colour-single
cluster with other partons. We define the probability of a parton 
leaks out from the confined volume as $w$. Thus, 

$$M^{(2-2)}(x_1,x_2\rightarrow x_3,x_4)[M^{(2-2)}(x_1,x_2\rightarrow x_3,x_4)]^*
\propto w ,\eqno(7.6)$$

and

$$M^{(2-1)}(x_1\rightarrow x_3,x_4)[M^{(2-3)}(x_2,x_1',x_2'\rightarrow x_3,x_4)]^* $$
$$+M^{(2-3)}(x_2,x_1',x_2'\rightarrow x_3,x_4)[M^{(2-1)}(x_1\rightarrow x_3,x_4)]^* 
\propto w^3, \eqno(7.7)$$
	
We can neglect the interference processes (7.7), because of the confinement 
condition $w<1$. In this case, we have another face of the evolution 
equation:
	
$$\frac{dG(x_B,Q^2)}{d\ln Q^2}$$
$$=(\frac{1}{RQ})^2
\int \sqrt{G(x_1,Q^2)G(x_2,Q^2)G(x_1+\Delta,Q^2)G(x_2-\Delta,Q^2)}$$
$$R_{GG}(x_1, x_2)R_{GG}(x_1+\Delta, x_2-\Delta)$$
$$\sum_i\kappa^{-1}_iP_{GG\rightarrow GG}^i(x_1,x_2,x_3,x_4,\Delta)
\delta(x_1+x_2-x_3-x_4)$$
$$[\delta(x_3-x_B)+\delta(x_4-x_B)-\frac{1}{2}\delta(x_1-x_B)-
\frac{1}{2}\delta(x_1+\Delta-
x_B)-\frac{1}{2}\delta(x_2-x_B)-\frac{1}{2}\delta(x_2-\Delta-x_B)]$$
$$dx_1dx_2dx_3dx_4d\Delta$$
$$-(\frac{1}{RQ})^2
\int \sqrt{G(x_1,Q^2)G(x_2,Q^2)G(x_1+\Delta,Q^2)G(x_2-\Delta ,Q^2)}$$
$$R_{GG}(x_1, x_2)R_{GG}(x_1+\Delta, x_2-\Delta)$$
$$\sum_i\kappa^{-1}_iP_{GG\rightarrow q\overline{q}}^i(x_1,x_2,x_3,x_4,\Delta)\delta(x_1+x_2-x_3-x_4)$$
$$[\frac{1}{2}\delta(x_1-x_B)+
\frac{1}{2}\delta(x_1+\Delta-
x_B)+\frac{1}{2}\delta(x_2-x_B)+\frac{1}{2}\delta(x_2-\Delta-x_B)]$$
$$dx_1dx_2dx_3dx_4d\Delta. 
\eqno(7.8)$$

and

$$\frac{dq(x_B,Q^2)}{d\ln Q^2}$$
$$=(\frac{1}{RQ})^2\int
\sqrt{G(x_1,Q^2)G(x_2,Q^2)G(x_1+\Delta,Q^2)G(x_2-\Delta,Q^2)}$$
$$R_{GG}(x_1, x_2)R_{GG}(x_1+\Delta, x_2-\Delta)$$
$$\sum_i\kappa^{-1}_iP_{GG\rightarrow q\overline{q}}^i(x_1,x_2,x_3,x_4,\Delta)\delta(x_1+x_2-x_3-x_4)
[\delta(x_3-x_B)+\delta(x_4-x_B)]dx_1dx_2dx_3dx_4d\Delta
\eqno(7.9)$$

Now the sign of the right-hand side of eq. (7.9) is positive.
This means that the shadowing in quark distribution is weaker than that
in gluon distribution.

	In principle, we can calculate the parton recombination functions
at order ${\cal O} (\alpha^2_s)$ for every parton flavors in the whole x 
region. However, in the majority of cases, the parton recombination 
is happened in the gluons with small x. For simplicity, we only consider 
the case where all partons are gluons with small x value in this work. 
We will discuss the recombination of partons in a general x range elsewhere. 
In this approximation, we can use the results of Mueller and Qiu 
in the calculations of the real process of fig. 1a in ref. [2]. 
Thus, the contributions of figs. 6a,b to $P^{(2-2)}_4$ 
are from t- and u-channels and as well as their interference-terms. 
One obtains

$$\sum_i\kappa^{-1}_iP_i^{(GG-GG)}\simeq\frac{\pi^3}{N^2-1}
(\frac{\alpha_s C_A}{\pi})^2\frac{x_4}{x_1^2}, \eqno(7.10)$$

where we assume that $x_1=x'_1=x_2=x'_2$. Eq. (7.10) evaluates to

$$2\int \sum_i\kappa^{-1}_iP_i^{(GG-GG)}\delta(x_3-x_B)\delta(2x_1-x_3-x_4)
dx_1dx_3dx_4$$
$$=\frac{4\pi^3}{N^2-1}
(\frac{\alpha_s C_A}{\pi})^2\int \frac{dx_1}{x_1}, \eqno(7.11)$$

which is the same as the result of ref. [2]. If we take the gluon correlation 
function [2] to be 

$$\frac{1}{R^2}G^{(2)}(x_1,x_1',x_2,x_2',Q^2)= 
\frac{9}{8\pi R^2}G(x_1,Q^2)G(x_1,Q^2), \eqno(7.12)$$ 

we obtain the following simplified evolution equations arising
from $GG\rightarrow GG$ in the small $x$ region:

$$\frac{dx_BG(x_B,Q^2)}{d\ln Q^2}
=-\frac{81}{16}(\frac{\alpha_s}{RQ})^2
\frac{[x_BG(x_B,Q^2)]^2}{x_B}$$
$$+\frac{81}{16}(\frac{\alpha_s}{RQ})^2x_B\int_{x_B/2}^{1/2}
\frac{[x_1G(x_1,Q^2)]^2}{x_1^3}dx_1$$
$$+P_{inter}\frac{81}{16}(\frac{\alpha_s}{RQ})^2
\frac{[x_BG(x_B,Q^2)]^2}{x_B}$$
$$-P_{inter}\frac{81}{8}(\frac{\alpha_s}{RQ})^2
x_B\int_{x_B}^{1/2}
\frac{[x_1G(x_1,Q^2)]^2}{x_1^3}dx_1.
\eqno(7.13)$$

Note that the modifications of fig. 5a-b are related to $G(x_B,Q^2)$
in eq. (7.13) but not to $x_BG(x_B,Q^2)$ according to eq. (5.1)
in our work. However, these real diagrams fig. 5a-b (or fig. 1a) in ref. [2] 
are regarded as the modifications in $x_BG(x_B,Q^2)$. 
Therefore, the equation (7.13) is different with the GLR equation in the 
dependencies of $x_1$ and $x_B$.

	In consequence, the new evolution equations if $P_{inter}=1$ are        

$$\frac{dx_BG(x_B,Q^2)}{d\ln Q^2}
=\frac{81}{16}(\frac{\alpha_s}{RQ})^2x_B\int_{x_B/2}^{1/2}
\frac{[x_1G(x_1,Q^2)]^2}{x_1^3}dx_1$$
$$-\frac{81}{8}(\frac{\alpha_s}{RQ})^2
x_B\int_{x_B}^{1/2}
\frac{[x_1G(x_1,Q^2)]^2}{x_1^3}dx_1.
\eqno(7.14)$$

Here, we have an extra conservation 

$$\frac{\int_0^1dx_Bx_BG(x_B,Q^2)}{d\ln Q^2}\equiv 0, \eqno(7.15)$$

since for any function $f(x_1)$ we have

$$\int_0^1dx_B\int_{x_B/2}^{1/2}f(x_1)dx_1-2\int_0^{1/2}dx_B
\int_{x_B}^{1/2}f(x_1)dx_1\equiv 0.
\eqno(7.16)$$
	
On the other hand, the new equation has following different form if
$P_{inter}=0$

$$\frac{dx_BG(x_B,Q^2)}{d\ln Q^2}$$
$$=-\frac{81}{16}(\frac{\alpha_s}{RQ})^2
\frac{[x_BG(x_B,Q^2)]^2}{x_B}
+\frac{81}{16}(\frac{\alpha_s}{RQ})^2
x_B\int_{x_B/2}^{1/2}
\frac{[x_1G(x_1,Q^2)]^2}{x_1^3}dx_1.
\eqno(7.17)$$

\begin{center}
\section{Discussions and conclusions}
\end{center}

	The following interesting components of the new evolution 
equation derived in this paper are highlighted:

\begin{enumerate}
\item   Through the derivations of sections 4 and 5, it seems 
there is an interesting ``cutting rule" in DIS:
The contributions of the cut diagrams in the sum (3.4) have the identical 
integral kernel with only the following different factors $R$:

$$R=(\pm)\times (1, \frac{1}{2})\times \delta(x_\beta-x_B). \eqno(8.1)$$

The various terms appearing in the cutting rule (8.1) can be described
in terms of the general structure of the cut diagrams $G_\gamma(N)$ in TOPT:

$$G_\gamma(N) =
\prod_{left-vertices}\frac{1}{\sum_{a'\in i+1} E_{a'}-\sum_{a\in i}E_a}
\prod_{right-vertices}\frac{1}{\sum_{b\in j} E_b-\sum_{b'\in j+1}E_{b'}}$$
$$\prod_{vertices-k}\delta(\sum_fx_f=0)
\prod_{states-c}\frac{1}{2E_c}
\prod_{loops-d}\frac{d^3{\bf k}_d}{(2\pi)^3}
\prod_{final-states-e}\frac{d^3{\bf k}_e}{(2\pi^3)^32E_e}N_G,
\eqno(8.2)$$

where $N_G$ is an overall numerator-and-symmetry factor
and is independent of the cut $\gamma$ [6]; $i$ and $i+1$ (or $j$ and $j+1$)
are the time-ordered lines on the left- (or right-) vertices;
$\delta(\sum_fx_f=0)$ is the conservation of longitudinal momentum at the
vertex.

\begin{enumerate}
\item The sign in the first factor of (8.1) is determined by
the energy deficits in (8.2).
For example, if a vertex pass through the cut line, the corresponding energy
deficit will change its sign since
$$\frac{1}{\sum_{a'\in i+1} E_{a'}-\sum_{a\in i}E_a}\rightarrow
\frac{1}{\sum_{a\in i} E_a-\sum_{a'\in i+1}E_{a'}}, \eqno(8.3)$$
as we have for example in (4.11).

\item The second factor takes a value of 1/2 if
the probe-vertex inserts in the initial line as shown in (5.11).

\item $\delta(x_\beta-x_B)$ is the direct result of the sum (3.4).

\item When the cut line moves its position, the contributions of the final
states in (8.2) will change the momentum-symbols,  
but don't change the structure of the intermediate state according to
the sum-condition I in (3.4). We also note that a virtual parton line has
4-dimensional integral and a real parton only a 3-dimensional
integral in the covariant perturbation theory; however, in TOPT,
since the virtual partons are replaced by
the effective real partons, the above mentioned differences are
contained in the energy deficits. In particular, when the cut line
moves to cutting line a $c$ from cutting a loop $d$-$d'$ in the process
of $c\rightarrow d'd \rightarrow e$, we have similar expressions
in (8.2) due to

$$\frac{1}{2E_c}\frac{d^3{\bf k}_{d'}}{(2\pi)^32E_{d'}}
\frac{d^3{\bf k}_d}{(2\pi)^32E_d}
\rightarrow
\frac{d^3{\bf k}_c}{(2\pi)^32E_c}
\frac{d^3{\bf k}_d}{(2\pi)^3}\frac{1}{2E_d}\frac{1}{2E_{d'}}.
\eqno(8.4)$$

Thus, as the cut line moves from cutting a loop to uncutting a loop,
the integral kernel has the same form. The difference lies only in the
cutting positions.

\item The cut line can also cut the nonperturbative matrix elements with
multi-initial partons. The reason is that the initial parton
line on the light-cone can be moved from one-side of the cut to another side
(see fig.6) [13, 14]. Thus, we can use the same correlation function to
represent the different hadronic parts and keep the same integral kernel.
Because of this important property, the sum (3.4) shall include more
complex cut diagrams when we study parton fusion or recombination.

\item Finally, the matrix can been factorized to obtain 
the probability explanation; the reason is that we used the bare 
probe-vertex approximation and the coefficient $C_q$ in (2.2) 
(or $C_\phi$ in (5.2)) is canceled in the calculations at this approximation.

\end{enumerate}

\item    In the DGLAP equation, there are IR-divergences 
when a final-state gluon becomes soft. These IR-divergences can be taken  
care of by the sum of figs. 4 according to eqs. (3.4) and (8.1).  
The divergences in (7.11) are canceled due to the symmetry at small x 
approximation [2]. However, a soft initial parton also may give rise to 
IR-divergences in the parton recombination process in a general case. 
We now show that such IR-divergences can be canceled by using the same 
method in the DGLAP equation.
	
	For example, take $x_2=0$ in fig. 10; this implies $p_2=0$. 
Since the unpolarized structure functions only involve contributions 
from terms with even-twist, we have  
$x'_2=0$ at $x_2=0$. The momentum of $l_L$ and $l_R$ are determined by 
the down-vertices since the energy is not conserved in the up-vertices
in the TOPT,
i.e., $l_L=[(x_2-x_4)P-\frac{l_{\perp}^2}{2x_4P}, l_{\perp}, (x_2-x_4)P]$ 
and $l_R=[(x_4-x_2')P+\frac{l_{\perp}^2}{2x_4P}, -l_{\perp}, (x_4-x_2')P]$. 
Thus, $x_4=0$ and $x_1=x_1'=x_3$. According to (5.14), we can find that  

$$\int f(x_1=x_B,x_2=0,x'_1=x_B,x'_2=0,Q^2)$$
$$[P_4^{(2-2)}(x_1=x_3,x_2=0,x'_1=x_3,x'_2=0,x_3,x_4=0)
\delta(x_3-x_B)dx_3$$
$$-\frac{1}{2}P_4^{(2-2)}(x_1,x_2=0,x'_1=x_1,x'_2=0,x_3=x_1,x_4=0)
\delta(x_1-x_B)dx_1$$
$$-\frac{1}{2}P_4^{(2-2)}(x_1=x_1',x_2=0,x'_1,x'_2=0,x_3=x_1',x_4=0)
\delta(x_1'-x_B)dx_1']=0.
\eqno(8.5)$$
	
Therefore, IR-divergences can be canceled point-by-point at the IR-pole.

\item Obviously, the new evolution equations (7.4), (7.5), (7.8) and 
(7.9) are different from the GLR equation [1,2]. It is interest that 
the properties and structure of the simplified low-$x$ form (7.14) is 
similar to a modified GLR equation that has been obtained earlier in ref. [15]. 
However, two equations really have a different theoretical basis. 
The GLR equation and its modified form are based upon the AGK cutting rules. 
They sum three kinds of diagrams: cutting two-ladders, one-ladder 
and zero-ladder, respectively. The first (see fig. 1a) is identical with 
our figs. 5a,b, however, the cut lines in the latter two figures (figs. 1b,c) 
break parton recombination and these processes should be inhibited.

\end{enumerate}

In conclusion, parton recombination via QCD evolution equation
is investigated using perturbative theory without the AGK cutting rules.
The contributions from different cut and interference diagrams are 
summed and infrared safeness and momentum conservation are established.
Time ordered perturbation theory is developed to establish the connections 
among different cut diagrams. As a consequence, a new nonlinear evolution
equation is derived on a different basis from the GLR equation.
Furthermore, this new evolution equation is more detailed in structure
than the GLR equation.

\vspace{0.3cm}

\noindent {\bf Acknowledgments}:

I would like to thank Jianwei Qiu for very helpful discussions and 
drawing my attention to reconsideration of the parton recombination in TOPT. 
I would also like to acknowledge D. Indumathi and Jianhong Ruan for useful 
comments. This work was supported by National Natural Science Foundation of 
China and `95-Pandeng' Plan of China.

\newpage
\noindent {\bf Figure Captions}

\noindent Fig. 1 Two expressions of the GLR equation based on the AGK
cutting rules: (left) cut vertex theory and (right) time-odered 
perturbation theory; where the cuts through (a) two-, (b) one- and 
(c) zero-ladders, respectively. 

\noindent Fig. 2 The amplitudes (a) $M^{(2-1)}_{p\gamma^*\rightarrow k'l'X}$,
(b) $M^{(2-2)}_{p\gamma^*\rightarrow k'l'X}$ and (c)
$M^{(2-3)}_{p\gamma^*\rightarrow k'l'X}$. The dark circles denote
the PQCD interaction with the correlation of the initial partons.

\noindent Fig. 3 Naive parton model of DIS.

\noindent Fig. 4 The leading order splitting processes in DIS.

\noindent Fig. 5 The diagrams contributing to the leading recombination order
from $\vert M^{(2-2)}_{p\gamma^*\rightarrow k'l'X}\vert ^2$.

\noindent Fig. 6 Identical hadronic parts in different cut graphs from 
refs.[13,14].

\noindent Fig. 7 The diagrams contributing to the leading recombination order
from 

$2M^{(2-1)}_{p\gamma^*\rightarrow k'l'X}[M^{(2-3)}_{p\gamma^*\rightarrow k'l'X}]^*$.

\noindent Fig. 8 A diagrammatic illustration of the relation between
$P^{(2-2)}_4$ and $P^{(1-3)}_4$.

\noindent Fig. 9 A diagrammatic illustration of the time
reversal invariance in $\gamma p \rightarrow \gamma p$.

\noindent Fig. 10 IR safeness in $GG\rightarrow GG$.

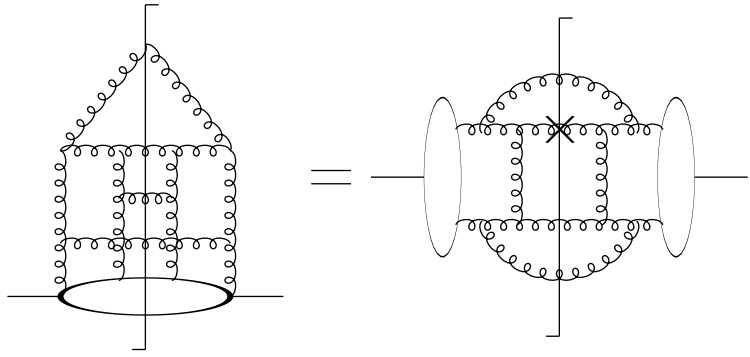
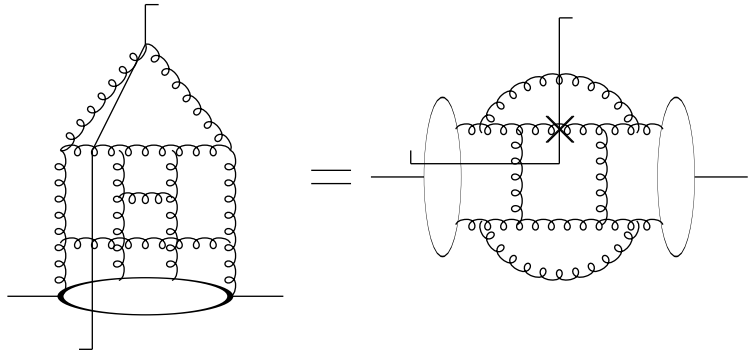
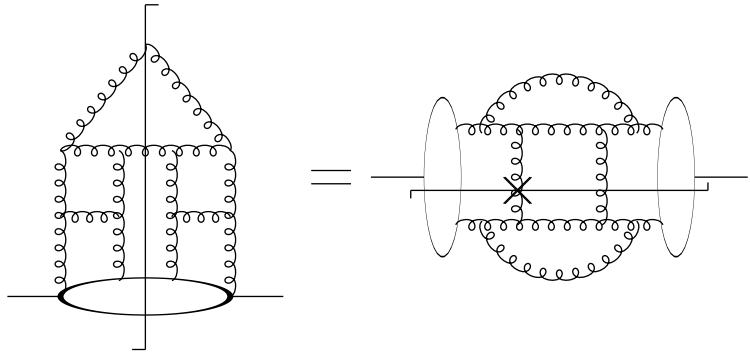
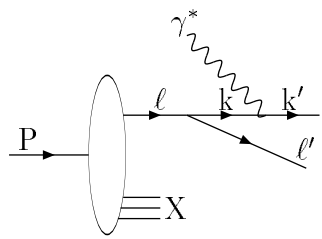
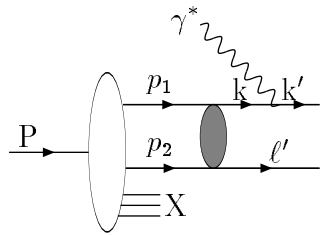
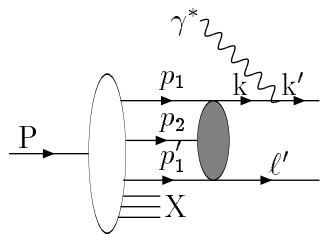
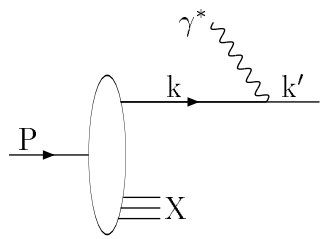
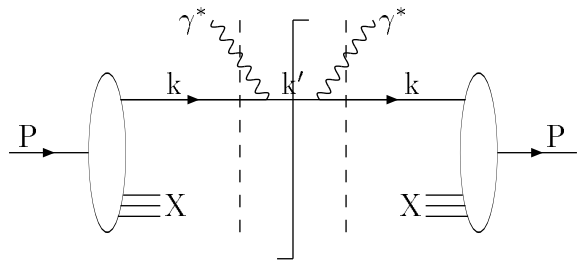
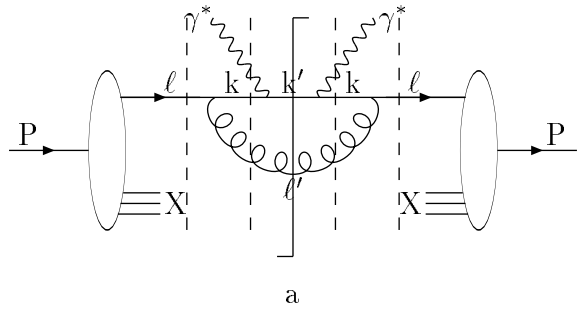
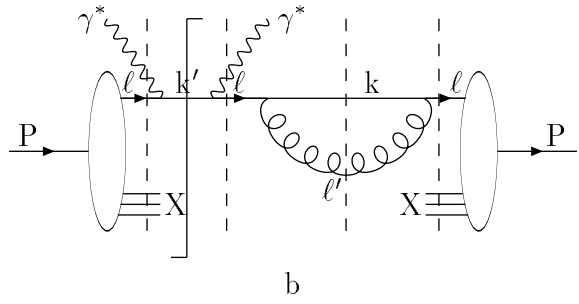
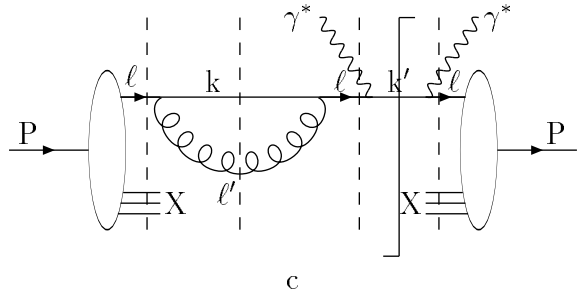
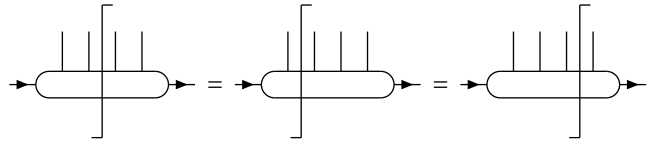
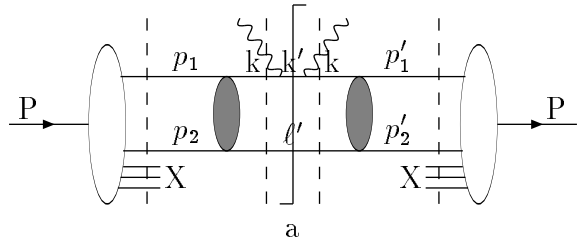
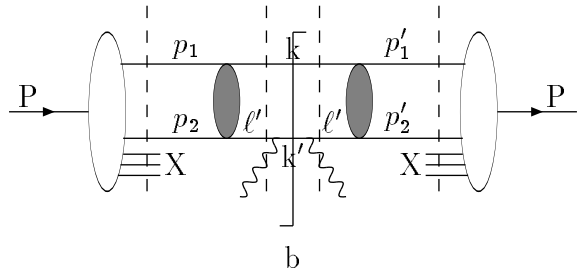
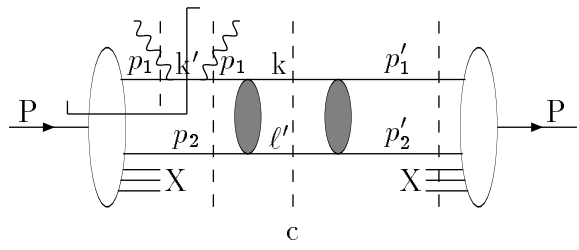
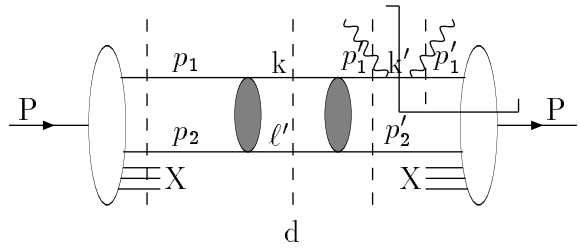
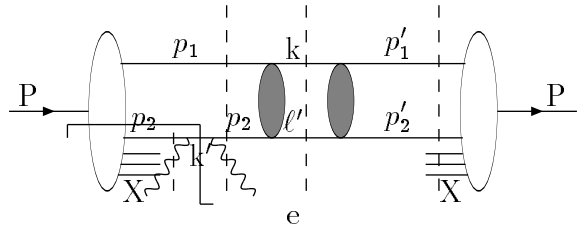
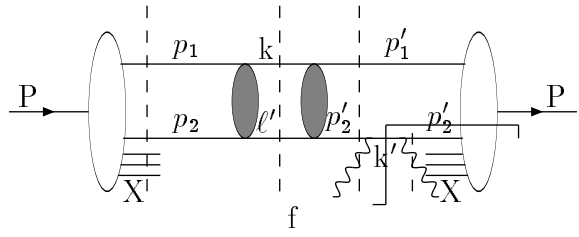
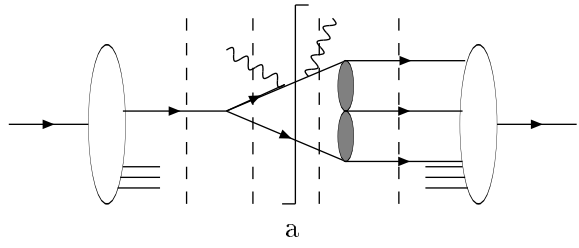
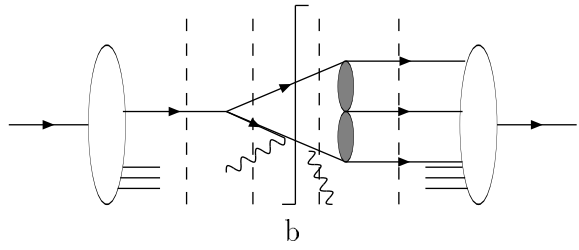
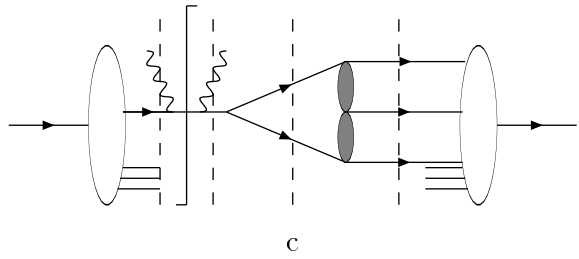
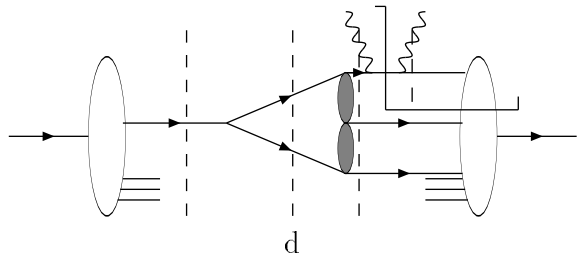
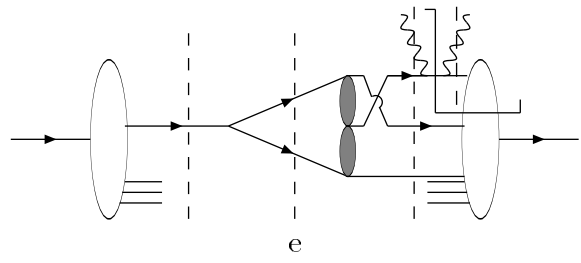
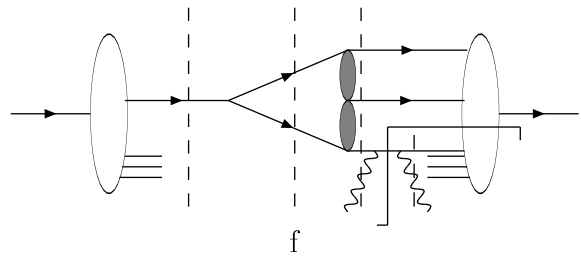
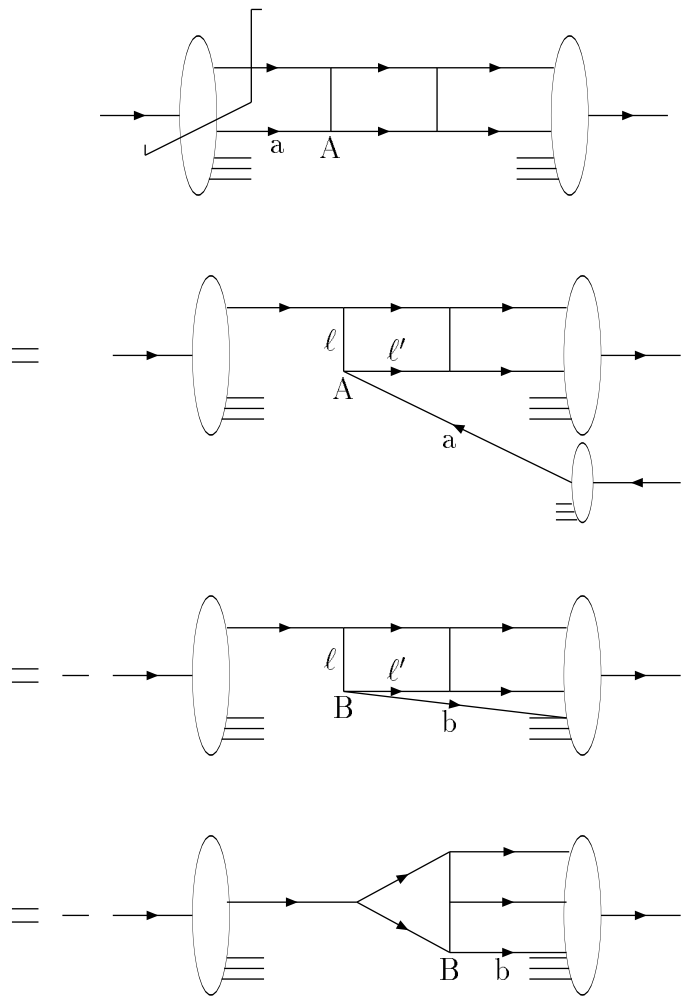
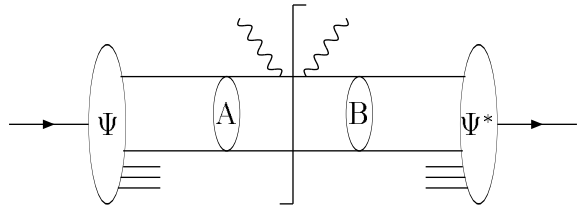
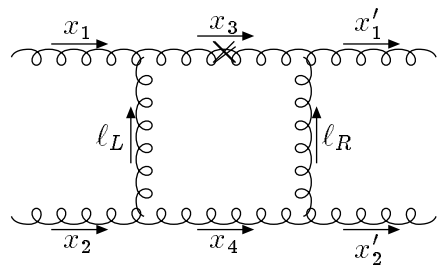
\end{document}